\documentclass[aps,prc,reprint,showkeys,showpacs,nofootinbib,superscriptaddress,floatfix]{revtex4-1}

\usepackage[english]{babel}
\usepackage{amsmath,amssymb}
\usepackage{bm}
\usepackage{graphicx}
\usepackage{braket}
\usepackage[colorinlistoftodos]{todonotes}
\usepackage[colorlinks=true,allcolors=blue]{hyperref}

\newcommand{\unedf}{\textsc{unedf}}
\newcommand{\pnfam}{\textsc{pnfam}}
\newcommand{\hfbtho}{\textsc{hfbtho}}

\providecommand{\jj}{\ensuremath{\mathbb{J}^2}}

\providecommand{\dd}{\ensuremath{\mathrm{d}}}
\providecommand{\vv}[1]{\ensuremath{\mathbf{#1}}}

\providecommand{\skop}{SkO$^\prime$}

\providecommand{\ccf}{\ensuremath{C_1^F}}

\newcommand{\opp}{\hat{\mathcal{O}}_p}
\newcommand{\opr}{\hat{\mathcal{O}}_r}
\newcommand{\oprsone}{\hat{\mathcal{O}}_{rs1}}

\begin{document}

\title{Finite Amplitude Method for Charge-Changing Transitions\\in Axially-Deformed Nuclei}
\date{\today}
\author{M.\ T.\ Mustonen}
\email{mika.t.mustonen@unc.edu}
\affiliation{Department of Physics and Astronomy, CB 3255, University of North Carolina, Chapel Hill, NC 27599-3255}
\author{T.\ Shafer}
\email{tshafer@physics.unc.edu}
\affiliation{Department of Physics and Astronomy, CB 3255, University of North Carolina, Chapel Hill, NC 27599-3255}
\author{Z.\ Zenginerler}
\email{zenginer@live.unc.edu}
\affiliation{Department of Physics, Faculty of Arts and Sciences, Sakarya University, 54100 Serdivan, Sakarya, Turkey}
\author{J.\ Engel}
\email{engelj@physics.unc.edu}
\affiliation{Department of Physics and Astronomy, CB 3255, University of North Carolina, Chapel Hill, NC 27599-3255}

\begin{abstract}
We describe and apply a version of the finite amplitude
method for obtaining the charge-changing nuclear response in the quasiparticle random phase approximation.  The method is suitable for calculating strength functions and
beta-decay rates, both allowed and forbidden, in axially-deformed open-shell
nuclei. We demonstrate the speed and versatility of the code through a
preliminary examination of the effects of tensor terms in Skyrme functionals on
beta decay in a set of spherical and deformed open-shell nuclei.  Like the
isoscalar pairing interaction, the tensor terms systematically increase allowed
beta-decay rates.  This finding generalizes previous work in semimagic nuclei
and points to the need for a comprehensive study of time-odd terms in nuclear
density functionals.  
\end{abstract}
\maketitle

% For figures -- \the\linewidth reports 246.0 pt as the width and 9.25pt as the caption text size (match axis label size to this??

\section{Introduction}

Beta decay is an important process at the intersection of nuclear physics,
astrophysics, and particle physics.  The rapid neutron-capture process ($r$
process) proceeds through neutron-rich nuclei, the beta-decay rates for which
determine final abundance distributions.  The significance of the reactor
neutrino anomaly for exotic new neutrino physics depends on forbidden
beta-decay rates in neutron-rich fission products \cite{Hayes13}.  In both
these cases, the important rates are difficult or impossible to measure; we
need to be able to calculate them instead.

The random-phase approximation (RPA) and its generalization, the quasiparticle
random-phase approximation (QRPA), nowadays typically used in conjunction with
Skyrme energy-density functionals (EDFs), are established tools for treating
nuclear excitations.  The matrix version of the charge-changing (or ``pn")
Skyrme QRPA has been applied with some success in spherical nuclei.  When the
rotational symmetry of the mean field is broken, however, the dimension of the
mean-field two-quasiparticle basis increases by orders of magnitude and the
QRPA matrix becomes too large to fit in the main memory of a typical computer
without aggressive truncation.  Even then, supercomputing is needed to solve the
equations.  We have constructed a deformed matrix Skyrme pnQRPA program
\cite{Mustonen13} from the code reported in Ref.\ \cite{Terasaki11} but cannot
use it in reasonable amounts of computing time.

The finite amplitude method (FAM) is a much more efficient scheme for finding
the linear response.  Ref.~\cite{Nakatsukasa07} first proposed the
method and Ref.\ \cite{Inakura09} quickly applied it to obtain the RPA
response in spherical and deformed nuclei.  Ref.~\cite{Avogadro11} generalized
the approach to the QRPA, and Ref.\ \cite{Stoitsov11} applied the
generalization to monopole transitions.  In this article, we further extend the
FAM to charge-changing QRPA transitions of arbitrary intrinsic angular momentum
projection $K$ in deformed nuclei.  We call the resulting approach the 
Skyrme proton-neutron finite amplitude method (pnFAM).  

To illustrate the method, we examine the effects of Skyrme's tensor terms on
beta-decay rates.  Minato and Bai~\cite{Minato13} observed that a tensor interaction
can reduce beta decay half-lives of magic and semi-magic nuclei considerably,
bringing them into closer accord with experiment.  If a similar reduction takes
place in deformed nuclei, it might make it impossible to include an isoscalar pairing
interaction without underpredicting half-lives. On the other hand, it might instead
allow a better-behaved isoscalar pairing interaction, one that depends less on mass than those
in use today.  After a
preliminary pnFAM analysis of the effects of tensor interaction in both
semi-magic and deformed nuclei, we assess the situation here.  This work will
serve as a stepping stone towards $r$-process studies in the rare-earth
region, evaluation of neutrino-capture rates, and a more data-rich
determination of the time-reversal (T) odd parts of energy-density
functionals.  

The rest of the article is organized as follows: Section~\ref{sec:theory}
lays out the form of the our Skyrme functionals and discusses the application
of the FAM to beta decay, Section~\ref{sec:method} presents our implementation
and consistency checks, and Section~\ref{sec:results}
uses the pnFAM to study the tensor interaction in a small set of open-shell and
deformed nuclei.  Section~\ref{sec:conclusions} is a conclusion.

\section{Theoretical Background}\label{sec:theory}

\subsection{Skyrme energy-density functional}\label{subsec:skyrme}

In the particle-hole channel we use the standard general Skyrme EDF, the
details of which may be found in many places, e.g.\ in
Refs.~\cite{Bender02,Perlinska04}.  In the notation of Ref.~\cite{Bender02},
the EDF takes the form
\begin{equation}\label{eq:skyrme-edf}
\mathcal E = \sum_{t=0,1} \sum_{t_3 = -t}^{+t}
\int \dd\vv r \, \boldsymbol{\left(}\mathcal{H}_{tt_3}^\mathrm{even}(\vv r) +
\mathcal{H}_{tt_3}^\mathrm{odd}(\vv r) \boldsymbol{\right)} \,,
\end{equation}
where
\begin{equation}\label{eq:h_even}
\begin{aligned}
\mathcal{H}_{tt_3}^\mathrm{even}(\vv r) &\equiv 
C_t^\rho[\rho_{00}] \rho_{tt_3}^2 + 
C_t^{\Delta\rho} \rho_{tt_3}\nabla^2\rho_{tt_3} \\ & + 
C_t^\tau \rho_{tt_3}\tau_{tt_3} + 
C_t^{J} \mathbb{J}^2_{tt_3} + 
C_t^{\rho\nabla J}\rho_{tt_3}\bm\nabla\cdot\vv J_{tt_3} 
\end{aligned}
\end{equation}
is bilinear in local time-even densities, and 
\begin{equation}\label{eq:h_odd}
\begin{aligned}
\mathcal{H}_{tt_3}^\mathrm{odd}(\vv r) &\equiv
C_t^s[\rho_{00}]\vv s_{tt_3}^2 +
C_t^{\Delta s}\vv s_{tt_3} \cdot \bm\nabla^2 \vv s_{tt_3} +
C_t^{j} \vv j_{tt_3}^2 \\ & + 
C_t^T \vv s_{tt_3} \cdot \vv T_{tt_3} + 
C_t^{s\nabla j} \vv s_{tt_3} \cdot \bm\nabla \times \vv j_{tt_3} \\ & +
C_t^F \vv s_{tt_3} \cdot \vv F_{tt_3} + 
C_t^{\nabla s} \left(\bm\nabla \cdot \vv s_{tt_3}\right)^2 
\end{aligned}
\end{equation}
is bilinear in time-odd local densities.  Only the coupling constants
$C_t^\rho[\rho_{00}]$ and $C_t^s[\rho_{00}]$ are allowed to be
density-dependent themselves, viz.:
\begin{equation}
\begin{aligned}
C_t^\rho[\rho_{00}] &= C_{t,0}^\rho + C_{t,\rho}^\rho \rho_{00}^{\sigma_\rho}\\
C_t^s[\rho_{00}] &= C_{t,0}^s + C_{t,\rho}^s \rho_{00}^{\sigma_s} \,,
\end{aligned}
\end{equation}
and even they depend only on the total density
\begin{equation}
\rho_{00}(\vv r)   = \sum_\sigma \sum_\tau \hat\rho(\vv r \sigma \tau,
\vv r \sigma \tau) = \rho_n(\vv r) + \rho_p(\vv r) \,.
\end{equation}

Our implementation of the pnFAM, through a code we call \mbox{\pnfam}, is self-consistent and so must be preceded by a Hartree-Fock-Bogoliubov (HFB)
calculation, for which we use the popular code \hfbtho\ \cite{hfbtho166,hfbtho200}.  Because the pnFAM treats only charge-changing
transitions and \textsc{hfbtho} does not allow proton-neutron mixing, our results
depend only on charge-changing densities [those with isospin indices $(t,t_3) =
(1,\pm 1)$]; the usual Coulomb and kinetic contributions to the total
energy in Eq.~\eqref{eq:skyrme-edf} are not necessary.

%\subsubsection{Particle-particle channel}

In the particle-particle (pairing) channel, we use a density-dependent interaction of the form
\begin{equation}\label{eq:vpp}
V_\mathrm{pp} = \left( V_0 \hat\Pi_{T=0} + V_1 \hat\Pi_{T=1} \right) \left( 1 - \alpha \frac{\rho_{00}(\vv r)}{\rho_c}  \right) \delta(\vv r),
\end{equation}
where $\rho_c = 0.16$~fm$^{-3}$ is the saturation density of nuclear matter and
$\alpha \in [0,1]$ controls the density-dependence.  This form is similar to that
allowed by \hfbtho. The $T=0$ pairing term, however, has no effect in the HFB
calculation as long as explicit proton-neutron mixing is forbidden.  The
pairing strength $V_0$ is thus unconstrained by the mean field and becomes a
free parameter in our subsequent pnFAM calculation.  On the other hand the
$T=1$ pairing, though important in the HFB calculation, has no dynamical
effect on Gamow-Teller transitions.  It does play a role for other multipoles, however, and we set its strength $V_1$ to the average
of the HFB proton and neutron pairing strengths (which \hfbtho\ allows to be
different), that is $V_1 = (V_p + V_n)/2$.

%\subsubsection{Time-odd coupling constants}

Although the coupling constants of Eqs.~\eqref{eq:h_even} and \eqref{eq:h_odd}
can be derived from the parameters that specify a Skyrme ``interaction"
(the $t$ and $x$ parameters) \cite{Perlinska04}, there need be no underlying interaction and the couplings of the time-odd part of the functional need not be connected with those of the time-even part.  Most Skyrme EDFs are fitted to
ground-state properties of spherical or axially-symmetric even-even nuclei,
which are independent of the time-odd functional.  Even if properties of odd
nuclei are included, time-odd densities and currents appear not to
contribute very much \cite{Schunck10}.  As a result, up to relations that follow from
gauge invariance, the time-odd couplings in the EDF picture are undetermined by
such fits.  In recent parameterizations, e.g.\ the \unedf\  \cite{UNEDF0,UNEDF1,UNEDF2} and SV \cite{Klupfel09} series of functionals, that fact is made explicit:  the time-odd coupling constants are either neglected completely or are constrained solely by gauge invariance.  

Some time-odd couplings can be profitably fit to the energies and strengths of Gamow-Teller resonances; see, e.g., Ref.\ \cite{Bender02} or Ref.\
\cite{Roca-Maza12}.  Here we will sometimes use the simple prescriptions of Ref.\ \cite{Bender02}.  The ability to treat charge-changing resonances
in deformed nuclei via the pnFAM should soon open the door to a better determination of the T-odd functional.

When tensor terms are included, there is an additional subtlety in the
time-even channel that determines mean-field properties in even-even nuclei.  The term involving the spin-current tensor $\mathbb J$ can actually be decomposed
into three tensor components \cite{Perlinska04}:
\begin{equation}\label{eq:tensor_decomposition}
C_t^J \jj \longrightarrow C_t^{J_0} J_t^2 + C_t^{J_1} \vv{J}^2_t + C_t^{J_2} \mathsf{J}^2_t
\end{equation}
Most groups (e.g.\ the authors of Ref.\  \cite{Lesinski07}) have thus far fit
the tensor terms solely in spherical nuclei.  There, the spin-orbit density
$\vv J$ is the only non-vanishing component of
Eq.~\eqref{eq:tensor_decomposition} \cite{Lesinski07}.  When spherical symmetry is broken, however, the contribution of the pseudotensor $\mathsf{J}$ is non-zero in the mean field and introduces another (undetermined) coupling constant.  In our calculations with \hfbtho, the coupling constant multiplying $\mathsf J$ is
restricted to obey the relation $C_t^{J_2} = 2C_t^{J_1} = C_t^J$.  As detailed in Ref.~\cite{Bender09}, this is only one of several options in deformed nuclei.  The choice means that we cannot require both that the
functional be gauge-invariant and that it have a non-vanishing \ccf.

\subsection{The Finite Amplitude Method}

Ref.\ \cite{Avogadro11} derives a form of the FAM that corresponds to the like-particle QRPA.  The formulation is general
enough, however, to cover the charge-changing case as well.  In the following
we discuss the special features of the FAM that follow from charge changing, i.e.\ from choosing an external field that transforms neutrons into protons (e.g. for $\beta^-$ decay).

Charge-changing transitions are generated by a weak external field $F(t)$, with (complex)
angular frequency $\omega$, of the form
\begin{equation}\label{eq:f1}
	F(t) = \eta (F e^{-i\omega t} + F^\dag e^{i\omega t}) \,,
\end{equation}
where $\eta$ is a small real parameter and $F$ is a one-body operator that
could depend on $\omega$ but in our application does not.  Transformed to the
quasiparticle basis, it has the form
\begin{equation}
\label{eq:f2}
F = \sum_{(\alpha,\beta)} ( F^{20}_{\alpha\beta} a^\dag_\alpha a^\dag_\beta +
F^{02}_{\alpha\beta} a_\beta a_\alpha) + \cdots \,,  
\end{equation}
where the ellipses refer to terms of the form $a^\dag_\alpha a_\beta$ that do
not contribute to the linear response.  The summation runs over every pair of
quasiparticle states in the basis, avoiding double-counting.

A one-body $\beta^-$ transition operator (Fermi, Gamow-Teller, or forbidden) can be written in a single-particle basis as 
\begin{equation}
F = \sum_{pn} f_{pn} c_p^\dag c_n \,,
\end{equation}
where the index $p$ runs over proton states and the index $n$ over neutron
states.  The $f_{pn}$ are the single-particle matrix elements of the transition
operator.  Here, unlike in the charge-conserving FAM, $F$ is non-Hermitian.
Without proton-neutron mixing in the static HFB solution, the Bogoliubov
transformation yields
\begin{subequations}
\begin{equation}
F^{20}_{\pi\nu} = \sum_{pn} U^*_{p\pi} f_{pn} V^*_{n\nu}\,,
\quad F^{20}_{\nu\pi} = 0 \,,
\end{equation}
and
\begin{equation}
F^{02}_{\pi\nu} = -\sum_{pn} V_{p\pi} f_{pn} U_{n\nu}\,,
\quad F^{02}_{\nu\pi} = 0 \,.
\end{equation}
\end{subequations}
where $\pi$ and $\nu$ label the proton and neutron quasiparticle states, and U and V are the usual Bogoliubov transformation matrices
\cite{ring-schuck}.

The weak external field $F$ induces weak time-dependent oscillations in the
quasiparticle annihilation operators (e.g.\ for neutrons),
\begin{equation}
\label{eq:annihilation-osc}
\delta a_\nu = \eta \sum_{\pi} a^\dag_\pi \left( X_{\pi\nu}(\omega)
e^{-i\omega t}+ Y_{\pi\nu}^*(\omega) e^{i\omega t}\right) \,.
\end{equation}
These oscillations in turn lead to oscillations in the charge-changing
density matrix elements $\rho_{pn}$ and $\rho_{np}$, the charge-changing pairing
tensors $\kappa_{pn}$ and $\kappa_{pn}^*$, and the resulting energy functional
$\mathcal{E}[\rho, \kappa, \kappa^*]$.  The single-particle Hamiltonian $h$ and
pairing potential $\Delta$ likewise acquire time-dependent pieces through the
relations
\begin{equation}\label{eq:functional_derivatives}
h_{ab} = \frac{\partial \mathcal{E}}{\partial \rho_{ba}}, \quad
\Delta_{ab} = \frac{\partial \mathcal{E}}{\partial \kappa^*_{ab}} \,,
\end{equation}
where $a$ is a proton index and $b$ a neutron index, or vice versa.  

The oscillations in all quantities occur with the same frequency $\omega$, and
the time-dependence, which is contained only in exponentials like those in Eq.\
(\ref{eq:f1}), can be factored out and removed.  The FAM then amounts to
solving the small-amplitude limit of the time-dependent HFB equation (with the
time dependence factored out).  The reason the procedure is so
efficient is the numerical computation of the derivatives of $h$ and $\Delta$
in the direction of the perturbation, i.e.\ with respect to $\eta$.  In our
pnFAM the differentiation is somewhat easier than in the like-particle case because the
charge-changing parts of $\rho$ and $\kappa$ vanish at the HFB minimum (since
our HFB doesn't mix protons with neutrons.)  That restriction, together with
the linear dependence of $h$ and $\Delta$ on the charge-changing densities for
all published Skyrme functionals, means that the pnFAM numerical derivatives
are independent of the parameter $\eta$.  In fact, our code doesn't reference
$\eta$ at all and we do not need to worry, as did the authors of Ref.\
\cite{Stoitsov11}, about choosing $\eta$ small enough so that terms of
$\mathcal{O}\left( \eta^2 \right)$ are negligible, but large enough to avoid
round-off errors.

When all is said and done, The pnFAM equations for the linear response become
\begin{equation}\label{eq:famequations}
\begin{cases}
(E_\pi + E_\nu - \omega)X_{\pi\nu}(F;\omega) + \delta H^{20}_{\pi\nu}(F;\omega) = -F^{20}_{\pi\nu} \\
(E_\pi + E_\nu + \omega)Y_{\pi\nu}(F;\omega) + \delta H^{02}_{\pi\nu}(F;\omega) = -F^{02}_{\pi\nu}
\end{cases} \,,
\end{equation}
where the $E$'s are the HFB quasiparticle energies and $H^{20}$ and $H^{02}$ are
the pieces of the frequency-dependent HFB Hamiltonian matrix, expressible in
terms of $h$ and $\Delta$, that multiply the quasiparticle pair-creation and
annihilation operators, as in Eq.\ (\ref{eq:f2}) \cite{Avogadro11}.  One can
go on from Eqs.\ \eqref{eq:famequations} to derive the traditional matrix-QRPA equations by expanding $\delta
H^{20}$ and $\delta H^{02}$ in $X$ and $Y$ (on which $H$ depends implicitly via the densities) and taking the limit of vanishing external field.  But the point of the FAM is to solve the nonlinear system of equations
(\ref{eq:famequations}) instead of constructing the traditional QRPA $A$ and
$B$ matrices.

After solving Eqs.\ (\ref{eq:famequations}) for the amplitudes
$X_{\pi\nu}(F;\omega)$ and $Y_{\pi\nu}(F;\omega)$, one can compute the strength
distribution for the operator $F$:
\begin{equation}\label{eq:transstrength}
\frac{dB(F, \omega)}{d\omega} = -\frac{1}{\pi} \operatorname{Im} S(F; \omega)
\,,
\end{equation}
where
\begin{equation}\label{eq:strfunc}\begin{split}
S(F;\omega) &= \sum_{\pi\nu} (F^{20*}_{\pi\nu} X_{\pi\nu}(F;\omega) + F^{02*}_{\pi\nu} Y_{\pi\nu}(F;\omega)) \\
&= -\sum_n \left( \frac{|\langle n|F|0 \rangle|^2}{\Omega_n - \omega} -
\frac{|\langle n|F^\dag|0 \rangle|^2}{\Omega_n + \omega} \right) \,.
\end{split}\end{equation}
The last form, combined with Eq.~\eqref{eq:transstrength} for complex frequency
$\omega = \Omega + i\gamma$, leads to
\begin{equation}\label{eq:lorentzian-smearing}
\frac{dB}{d\omega} \rightarrow \frac{\gamma}{2\pi} \sum_n \left( \frac{|\langle n|F|0
\rangle|^2}{(\Omega_n - \Omega)^2 + \gamma^2} - \frac{|\langle n|F^\dag|0
\rangle|^2}{(\Omega_n + \Omega)^2 + \gamma^2} \right) \, ,
\end{equation}
which shows that the FAM transition strength a distance $\gamma$ above the real
axis is just the QRPA strength function smeared with a Lorentzian of width
$\gamma$.  The last three equations imply the symmetry 
\begin{equation}
\label{eq:s-symmetry}
S(F; \omega) = -S^*(F; \omega^*) \,.
\end{equation}

To evaluate non-unique forbidden decay rates we will need to take into account the
interference between distinct transition operators, called $F$ and $G$ here for
simplicity.  Such terms have the form
\begin{equation}\label{eq:crossterms}
   \chi(F,G;\omega) = \sum_n \left( \frac{\langle n|F^\dag|0\rangle \langle 0|G|n \rangle}{\Omega_n + \omega} - \frac{\langle n|F|0\rangle \langle 0|G^\dag|n \rangle}{\Omega_n - \omega} \right).
\end{equation}
We compute them by using, e.g., the operator $F$ to generate the response
and then calculating the effect on the quantity represented by $G$:
\begin{equation}
\label{eq:crossterm}
\chi(F,G;\omega) = \sum_{\pi\nu} \left[ G^{20*}_{\pi\nu} X_{\pi\nu}(F;\omega) +
G^{02*}_{\pi\nu} Y_{\pi\nu}(F;\omega) \right]\,.
\end{equation}

In deformed nuclei, all the results above are in the intrinsic frame, where
angular momentum is not conserved.  The symmetry must be restored, at least
approximately, and the crudest way to do so is by treating the intrinsic state like the particle in the particle-rotor model \cite{ring-schuck}.  In that picture every intrinsic state corresponds to the lowest state in a rotational band, and has a rotational energy
\begin{equation}
\label{eq:cranking}
E_\textrm{lab}(J) = E_\textrm{int} + \frac{J(J+1)}{2\mathcal{I}} \,,
\end{equation}
where $\mathcal{I}$ is the moment of inertia of the nucleus.  We use the HFB
version of the Belyaev formula \cite{ring-schuck} (see the Appendix for
details) to approximate $\mathcal{I}$:
\begin{equation}
\label{eq:rotcorrection}
\mathcal{I} = \sum_{\alpha\beta} \frac{|(U^\dag J_x V^* - V^\dag J_x
U^*)_{\alpha\beta}|^2}{E_\alpha + E_\beta} \,.
\end{equation}
Here $\alpha$ and $\beta$ label quasiparticle states of the same particle type
(proton or neutron).  The energy shifts are typically only tens of keV, but their effects in beta-decay rates are magnified by the phase-space integrals and can be non-negligible (see Sec.~\ref{sec:results}).

\subsection{Beta-decay half-lives}

In this work we consider both allowed and first-forbidden beta decay.
Expressions for the relevant impulse-approximation operators were worked out some time ago, e.g.\ in Refs.\
\cite{Behrens82,Schopper66,Suhonen93}.  In this section we restrict ourselves
to allowed decay; the more complicated expressions for the first-forbidden
decay can be found in Appendix~\ref{app:forbidden-formulas}.

The total allowed decay rate is proportional to the sum of individual
transition strengths $B_i$ to all energetically allowed states $i$ in the
daughter nucleus, weighted by phase-space integrals:
\begin{equation}\label{eq:decayrate}
   \lambda = \frac{\ln 2}{\kappa} \sum_i f(E_i) B_i,
\end{equation}
where the constant $\kappa = (6147.0 \pm 2.4)$~s comes from superallowed decay
\cite{Hardy05}.  The phase-space integral, containing the details of
final-state lepton kinematics, is \cite{Behrens82}
\begin{equation}
\label{eq:fermiintegral}
f(E_0) = \int_{1}^{W_0} \dd W\, p \, W \left(W_0-W\right)^2 L_0 F_0(Z, W)\,,
\end{equation}
where $Z$ is the charge of the daughter nucleus, $W_0 = E_0 / (m_e c^2)$, $W$ is the electron
energy in units of electron mass, $p \equiv \sqrt{W^2 - 1}$ is the electron
momentum, and $F_0(Z, W)$ is one of the (generalized) Fermi functions \cite{Behrens82}
\begin{equation}
\label{eq:fermifunction}
\begin{gathered}
F_{k_e}(Z, W) = \left[\,k_e(2k_e-1)!!\,\right]^{2} 4^{k_e}
(2pR)^{2(\gamma_{k_e}-k_e)} \\ 
\quad \times \exp(\pi y)
\frac{\left|\Gamma\left(\gamma_{k_e}+iy\right)\right|^{2}}
	{\left[\,\Gamma\left(2\gamma_{k_e}+1\right)\,\right]^{2}} \,.
\end{gathered}
\end{equation}
Here $k_e$ is related to the orbital angular momentum of the emitted electron (see, e.g., Ref.~\cite{Behrens82} for the definition), $\gamma_{k_e} = \sqrt{k_e^2 - (\alpha Z)^2}$, $y=\alpha Z W/p$, and $R$ is the nuclear radius.  (The
Primakoff-Rosen approximation to this expression \cite{Primakoff59} is often
used for computing allowed decay but we retain the more general form, which 
also applies to forbidden beta decay.)  The Coulomb function $L_0$ is
\begin{equation}
\label{eq:coulomb}
L_0 \approx \frac{1}{2} (1+\gamma_1) \,.
\end{equation}

To use these expressions we need the energies in the final nucleus with respect
to the ground state of the initial nucleus.  We take our
ground-state-to-ground-state $Q$ value from the approximation in Ref.\
\cite{Engel99},
\begin{equation}\label{eq:qvalue}
   Q = \lambda_n - \lambda_p + \Delta M_\mathrm{n-H} - E_\mathrm{g.s.}\,,
\end{equation}
where $\lambda_p$ and $\lambda_n$ are the proton and neutron Fermi energies
from the HFB solution, $\Delta M_\mathrm{n-H} = 0.78227$~MeV is the mass
difference between the neutron and hydrogen atom, and the ground-state energy is taken to be the sum of the lowest proton and neutron quasiparticle energies:
\begin{equation}\label{eq:egs1}
E_\mathrm{g.s.} \approx E_\textrm{p, lowest} + E_\textrm{n, lowest} \,.
\end{equation}
%In nuclei where pairing collapses for either protons or neutrons, we take
%care to choose the proton particle state or the neutron hole state with the
%lowest energy.  
One virtue of the approximation in Eqs.\ (\ref{eq:qvalue}) and
(\ref{eq:egs1}) is that the independent-quasiparticle approximation to the
ground-state energy cancels out in calculations of beta-decay lifetimes
\cite{Engel99}. We use the approximation for the ground-state energy only when studying strength distributions.

How do we actually evaluate lifetimes from the pnFAM response?
Eq.~\eqref{eq:strfunc} implies that the transition strength of the operator $F$ between the QRPA state with energy $\Omega_n > 0$ and the initial ground state is the residue of the function $S(F)$ at that energy, 
\begin{equation}
B_n(F) = |\langle n| F| 0\rangle|^2 = \operatorname{Res}[S(F),\Omega_n] \,,
\end{equation}
and Eq.~\eqref{eq:crossterms} that the cross terms contributing to forbidden decay rates are
\begin{equation}
\langle n| F| 0\rangle \langle n| G | 0\rangle^* =
\operatorname{Res}[\chi(F,G),\Omega_n] \,.
\end{equation}
The connection to residues allows us to represent beta-decay rates as contour integrals of the pnFAM response in the complex-frequency plane.  

The representation is complicated a little by the fact that the phase-space
integrals \eqref{eq:phasespaceints} are not analytic functions.  But we can
replace the phase-space integrals with other functions that are analytic, at
least inside the contour, and that coincide with the phase-space integrals at the
poles of the strength function that contribute to the integral.  A
high-order polynomial of the form
\begin{equation}\label{eq:psi-poly}
f_\mathrm{poly}(\omega) = \sum_{n=0}^N a_n \left( \frac{\omega_\mathrm{max}-\omega}{m_ec^2} \right)^n
\,,
\end{equation}
fitted to the phase-space integral on the real axis, serves our purpose.
While we do not know the exact locations of the poles of the strength function,
we do know they lie on the positive real axis (and that mirrored, unphysical poles
lie on the negative real axis).

\begin{figure*}
\begin{center}\includegraphics[width=15cm]{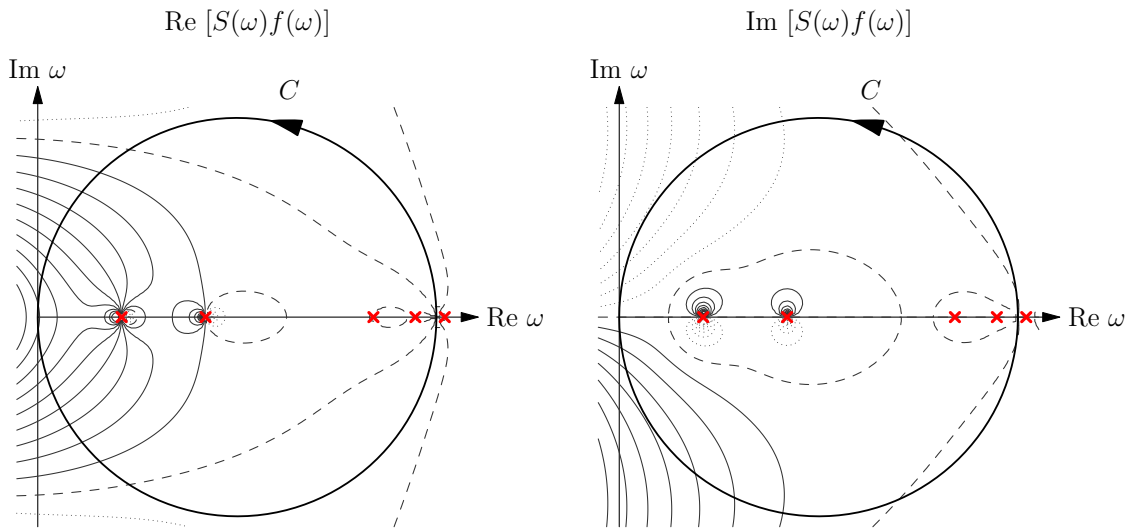}\end{center}
\caption{(Color online.) Schematic representation of the integration contour $C$ used to
evaluate beta-decay rates.  Only the poles of the strength function below the
endpoint energy $\omega_\text{max}$ contribute to the decay rate.}
\label{fig:contour}
\end{figure*}

With the polynomials, we can cast the equations for beta-decay rates in a form that captures the contributions of all the individual excited states in a contour that encloses them.  The Gamow-Teller part of the rate takes the form 
\begin{equation}
\label{eq:rate-integral}
\begin{split}
\lambda_{1^+} =& \frac{\ln 2}{\kappa} \sum_n f(\Omega_n) B_n^\textrm{(GT)} \\
\approx & -\frac{\ln 2}{\kappa} \sum_n f_\textrm{poly} (\Omega_n) \operatorname{Res}
[S(\sigma\tau_-),\Omega_n] \\
=& -\frac{\ln 2}{\kappa} \sum_n \operatorname{Res} [f_\textrm{poly}
S(\sigma\tau_-),\Omega_n] \\
=& -\frac{\ln 2}{\kappa} \frac{1}{2\pi i} \oint_C \dd\omega \: f_\textrm{poly} (\omega)
S(\sigma\tau_-;\omega),
\end{split}
\end{equation}
where the contour $C$ encloses the same poles $n$ that are initially summed
over.  A practical choice for the contour is a circle
\begin{equation}
\label{eq:contour}
 \omega(t) = \frac{\omega_\mathrm{max}}{2} (1 + e^{it}) \,,
\end{equation}
crossing the real axis at
the origin and at the maximum energy
\begin{equation}
\omega_\text{max} = Q + E_\textrm{g.s.} = \lambda_n - \lambda_p + \Delta M_\mathrm{n-H}
\,.
\end{equation}
Figure~\ref{fig:contour} displays such a contour schematically.  A circular contour allows the use of the symmetry in
Eq.\ \eqref{eq:s-symmetry} to halve the number of pnFAM computations.

The analog of Eq.\ \eqref{eq:rate-integral} for first-forbidden beta decay is
lengthy and presented in the Appendix~\ref{app:forbidden-formulas}.

\section{Computational Method and Tests}\label{sec:method}

As mentioned, our method begins with the use of the code \hfbtho\
\cite{hfbtho166,hfbtho200} to carry out an axially-deformed HFB calculation.
In our tests 16 harmonic oscillator shells are enough to allow the low-energy
strength functions to converge, and we adopt that number for all half-life
calculations.

\begin{figure}
\begin{center}\includegraphics{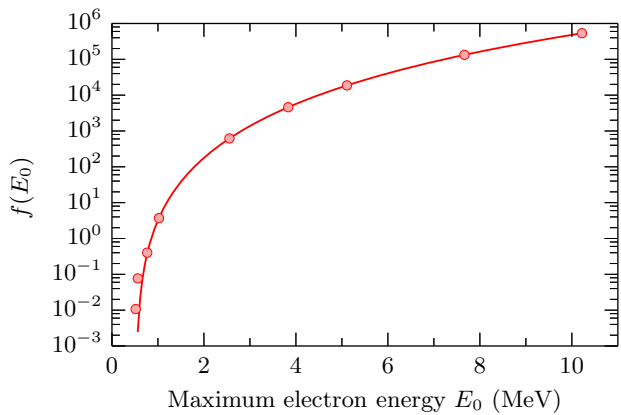}\end{center}
\caption{(Color online.) A 14th-order polynomial approximation to the phase space integral
$f(E_0)$ (Eq.~\eqref{eq:fermiintegral}) for the beta decay of $^{148}$Ba.  The
solid line is calculated with the exact Fermi function $F_0$ (Eq.~\eqref{eq:fermifunction}) and
the points correspond to the polynomial approximation \eqref{eq:psi-poly}.}
\label{fig:polyfit}
\end{figure}

\begin{figure}
\begin{center}\includegraphics[width=8.4cm]{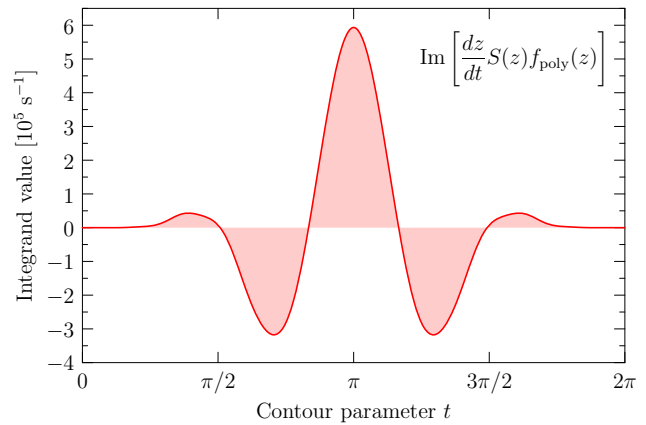}\end{center}
\caption{(Color online.) The imaginary part of the integrand in Eq.~\eqref{eq:rate-integral}
that determines the $K=0$ allowed decay rate of $^{142}$Ba, with SkO and the
tensor interaction.  The integrand behaves well enough to allow simple
quadrature.  The origin in the complex plane corresponds to $t=\pi$ (see text).}
\label{fig:contourint}
\end{figure}

Our contour integration requires a reasonably accurate polynomial approximation
to the Fermi integrals in Eq.\ \eqref{eq:phasespaceints}.
Figure~\ref{fig:polyfit} illustrates the quality of our fit to the
allowed-decay Fermi integral.  In practice, a 10th-order polynomial of
the form \eqref{eq:psi-poly} is more than sufficient.  Another requirement is  that the
integrands are smooth enough to allow numerical quadrature.
Figure~\ref{fig:contourint} demonstrates that that is the case, displaying a
typical integrand as a function of the curve parameter $t$ in Eq.\ \eqref{eq:contour}.
The integrand is indeed smooth enough to treat with conventional quadrature; we use the compound Simpson's $3/8$ rule. 

\begin{figure}
\begin{center}\includegraphics{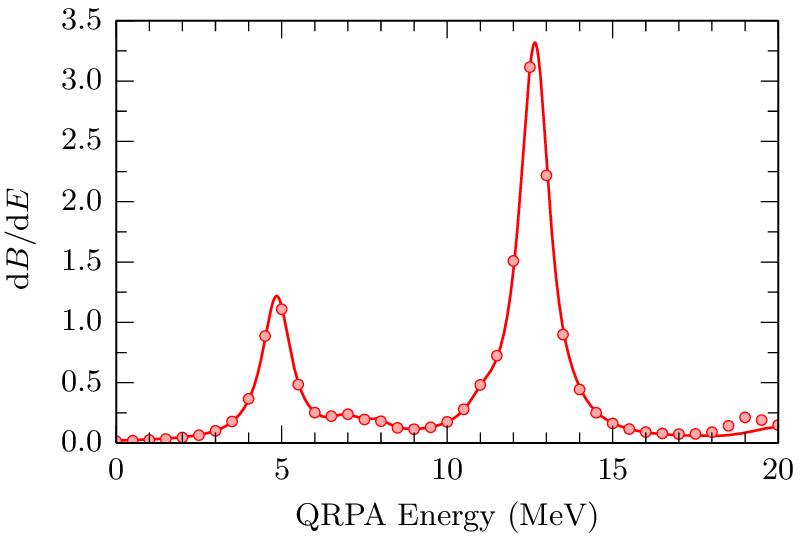}\end{center}
\caption{(Color online.) Comparison of the pnFAM Gamow-Teller strength function (points) in the deformed nucleus $^{22}$Ne
with the same function from the matrix QRPA (lines), smeared with a Lorentzian.  We use the Skyrme functional SkM* without
including \jj\ terms or pairing in the QRPA.}
\label{fig:matrix-comparison}
\end{figure}

To test the \pnfam\ solver itself, we compare in
Fig.\ \ref{fig:matrix-comparison} the pnFAM Gamow-Teller transition strength
function in the deformed isotope $^{22}$Ne with that produced by the traditional matrix-QRPA code used in Ref.~\cite{Mustonen13}.  The matrix code uses the Vanderbilt HFB solver \cite{Teran03} as its starting point.  The slight differences between the two strength functions are due to similarly slight differences in the HFB solutions, which in turn stem from different single-particle bases and truncation schemes. 

\begin{figure}
\begin{center}\includegraphics[width=8.4cm]{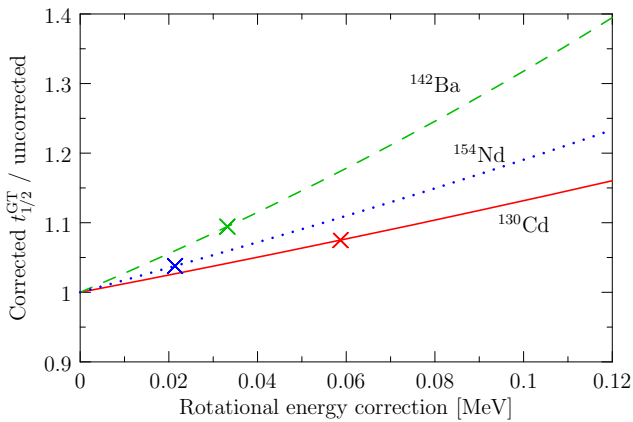}\end{center}
\caption{(Color online.) Partial Gamow-Teller half-lives in several nuclei as a function of
rotational energy correction, normalized to the uncorrected values.  The marker
on each curve indicates the correction obtained from
Eq.~\eqref{eq:rotcorrection}.}
\label{fig:rotcorr-sensitivity}
\end{figure}

Finally, we turn to our prescription for the nuclear moment of inertia.  The approximation in Eq.\ \ref{eq:rotcorrection} appears to yield systematically higher values than does experiment,
indicating that almost none of the nuclei we examine below are as rigid as the
straightforward extension of the Beliaev formula predicts.   (We can implement a better approximation that takes into account RPA correlations --- the Thouless-Valatin prescription \cite{Thouless62}  --- once a like-particle FAM for general $K$ exists.)  To assess the sensitivity of the
Gamow-Teller half-life to the rotational energy correction, we use the SkO functional detailed in the next section to calculate half-lives in
a few test nuclei (Fig.\ \ref{fig:rotcorr-sensitivity}).  Although the strong dependence of the phase-space integral on the energy released in the decay can make a correction of the order of ten percent to the half-life, that error is still at most comparable to the error in the calculated $Q$ value.  The accuracy of the generalized Beliaev moment of inertia is therefore good enough for use with present-day energy functionals. 

\section{Results and Discussion}\label{sec:results}

Recent work \cite{Bai09a,Bai09b,Bai11,Minato13} on semi-magic nuclei in the
spherical QRPA indicates that tensor terms in Skyrme EDFs have
significant effects on beta-decay rates.  Here we explore the issue in open
shell nuclei, both spherical and deformed.  We choose a set of isotopes for which both beta-decay rates and the allowed contribution to those
rates have been measured.  To make contact with Ref.~\cite{Minato13} we use the
same underlying SkO functional, with the same additional tensor piece
(i.e.\ the interaction parameters $t_e = 184.567\textrm{ MeV fm}^5$, $t_o = -108.567\textrm{ MeV fm}^5$, in the notation of Ref.\ \cite{Perlinska04}).  We also adopt the Ref.\ \cite{Minato13} procedure of breaking self-consistency by omitting
the central $J^2$ terms from the HFB calculation while including them in the
QRPA.  Unlike Ref.\ \cite{Minato13}, however, we include the rotational energy
correction and we approximate the ground-state energy by the sum of the lowest
proton and neutron quasiparticle energies.  These differences in procedure have
small effects on the $Q$ value and half-life (via the phase space available
to emitted leptons).  The last difference with Ref.\ \cite{Minato13}:  we use the quenched value $g_A=1.0$ rather than 1.27 for
the axial-vector coupling constant. 

We fit the constants in the isovector pairing interaction Eq.\ \eqref{eq:vpp}
to a three-point interpolation of measured separation energies.  The SkO
functional with this pairing interaction reproduces $Q$ values well, both with
and without tensor terms.

\begin{figure}
\begin{center}\includegraphics[width=8.5cm]{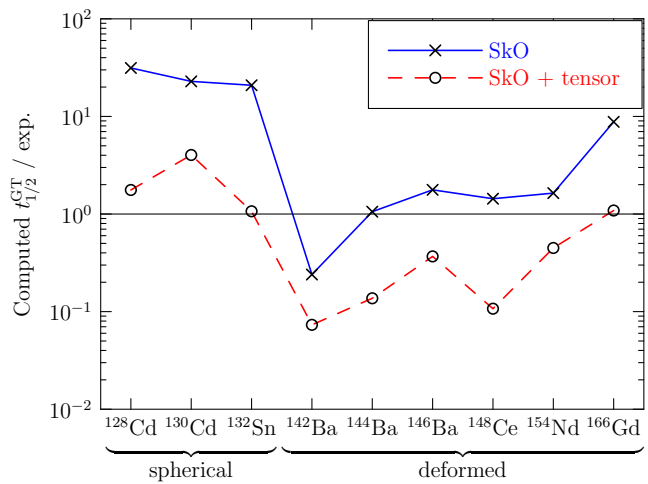}\end{center}
\caption{(Color online.) The ratio of computed and experimental partial Gamow-Teller
half-lives.  Two functionals are used:  the bare SkO functional and the SkO
functional with an added tensor piece.  Isoscalar pairing is absent here.  }
\label{fig:halflives}
\end{figure}

Fig.\ \ref{fig:halflives} shows the ratios of computed and experimental partial
Gamow-Teller half-lives for our set of nuclei.  The tensor interaction
systematically reduces the half-lives, as in the magic and semi-magic nuclei
examined by Ref.\ \cite{Minato13}.  In our spherical nuclei, with or without
open shells, the agreement with experiment improves dramatically.  In the
deformed isotopes, however, the half-lives with SkO tend to be quite low even
without the tensor terms, which actually make the half-lives too short.  The
situation is thus more complicated than it seems when restricted to spherical
systems.

\begin{figure*}
\begin{center}\includegraphics[width=17.8cm]{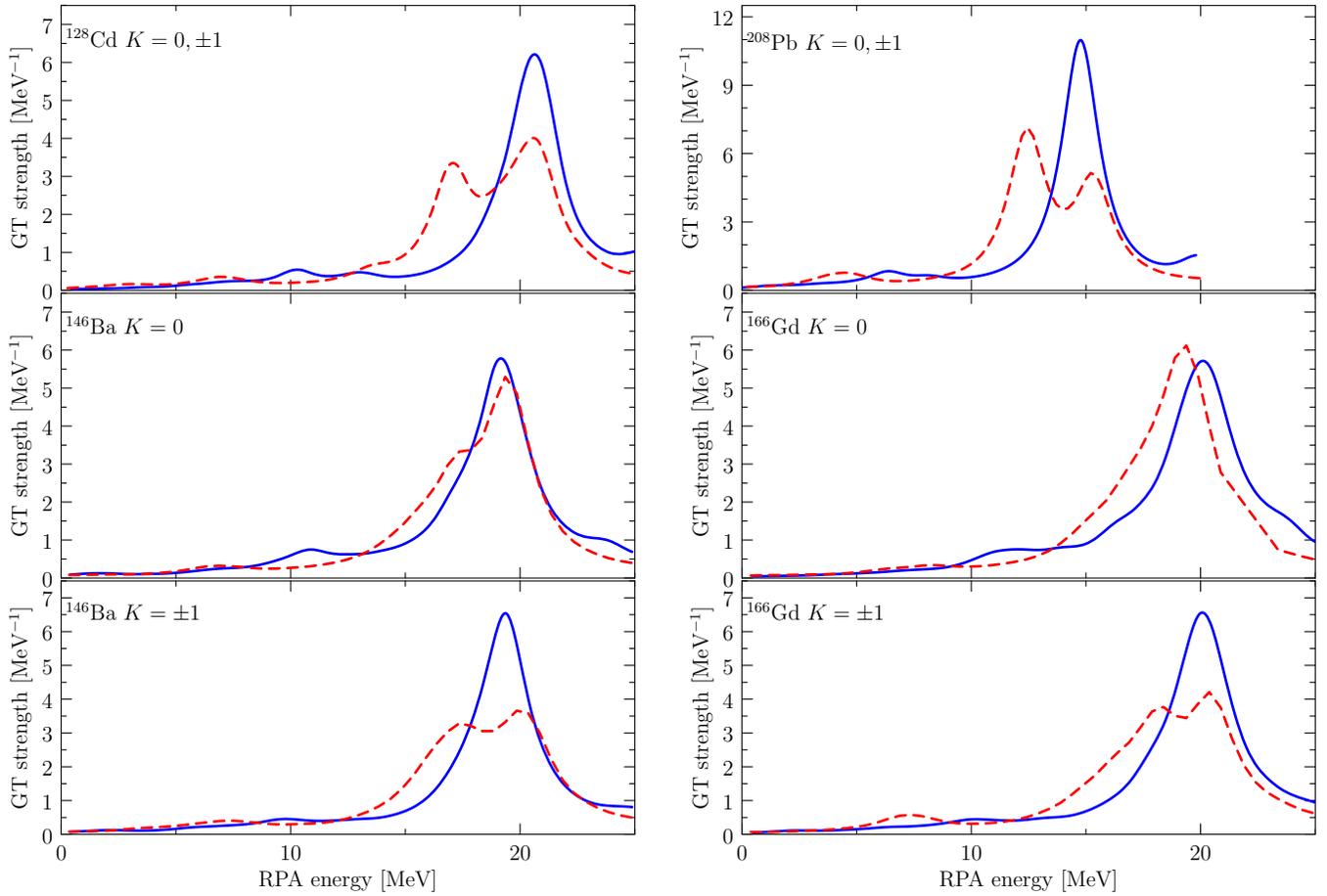}\end{center}
\caption{(Color online.) Gamow-Teller strength functions in several isotopes.
The solid (blue) lines represent the strength without tensor terms and the dashed (red) lines the strength with those terms.}
\label{fig:strfunction}
\end{figure*}

Figure~\ref{fig:strfunction} shows the effects of the tensor interaction in
more detail, in four isotopes.  The new terms pull Gamow-Teller strength down in energy in
each case, and smear the resonances.  The movement of strength to lower
energies explains the decrease in half-life; the lower-energy strength means
more phase space for leptons and an increased rate.

How many of the features in Figs.\ \ref{fig:halflives} and
\ref{fig:strfunction} are due to the violation of self-consistency between the
HFB and QRPA calculations?  How many are due to the limited set of nuclei that
we examine?  To the restriction to allowed decay?  To the simple addition of a
tensor interaction without any attempt to refit data?  We can't address these
questions fully here, but can make a start.  We now investigate a slightly larger set of
nuclei (that overlaps our original set) with a fully self-consistent
calculation that includes first-forbidden contributions to the rate.  We
choose as a starting point the functional \skop\ \cite{Reinhard99}, which
reproduces experimental $Q$ values as well as SkO and does a good job on
beta-decay rates in semi-magic isotopes \cite{Engel99}.  We adjust the
time-odd part of the functional, setting $C^{\Delta s}_1 = 0$ to avoid
instability (observed, e.g., in \cite{Schunck10}) and $C^s_1 = 159 \textrm{ MeV
fm}^3$ to reproduce the Gamow-Teller resonance energy in $^{208}$Pb.  We leave the
other coupling constants untouched.  When we include the tensor interaction we
use the values implied by the Skyrme $t$ and $x$ parameters; the relations between these parameters and the $C$'s are given, e.g., in Ref.\ \cite{Perlinska04}. All this is the same prescription for the time-odd terms that was found practical (without tensor terms) in Ref.\ \cite{Bender02}.

\begin{figure}
\begin{center}\includegraphics[width=8.5cm]{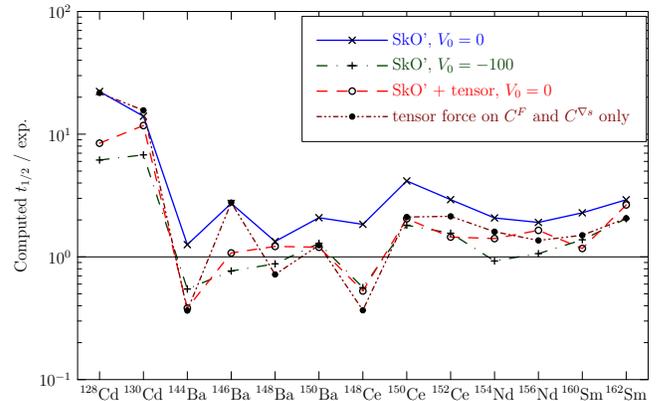}\end{center}
\caption{(Color online.) Ratio of calculated to experimental half-lives with the
functional \skop and first-forbidden contributions included.  The bare \skop\
half-lives (solid blue line) are systematically longer than experiment and can
be reduced by introducing either isoscalar pairing (dashed green line), a full
tensor interaction (dashed red line), or only the time-odd components
associated with such an interaction (dotted line).}
\label{fig:halflives-skop}
\end{figure}

Fig.\ \ref{fig:halflives-skop} shows some of the results.  Without the tensor
interaction it is possible to roughly reproduce the half-lives through an appropriate strength for the isoscalar pairing interaction in Eq.\ \ref{eq:vpp} (about 60\%
of isovector interaction strength); in the analysis leading to Figs.\ \ref{fig:halflives} and 
\ref{fig:strfunction} we did not include isoscalar pairing, which has been
the most convenient remedy for many of the QRPA's deficiencies.  Ref.\
\cite{Minato13} suggests that the tensor interaction can obviate strong
isoscalar pairing.  To begin to test this idea, we add the same tensor terms we
used with the SkO functional.  There is no particular justification for this
choice other than its successes with SkO and the lack of any work on tensor
interactions in conjunction with \skop.  Yet, as Figure~\ref{fig:halflives-skop} shows, these tensor
terms lower the half-lives in very much the same way as isoscalar pairing.

As mentioned above, however, the simple addition of a tensor interaction spoils the
functional's ability to reproduce data.  We have compensated for the problem in
the time-odd channel by readjusting $C^s_1$, but have done nothing to repair
the time-even channel, the original parameters of which were obtained through
careful fits to energies, radii, etc.  We therefore look at what happens when
we leave the time-even part of \skop\ alone, adding tensor terms to the time-odd part only.  
To make the changes truly minimal, we allow the tensor interaction to alter only the
two time-odd coupling constants $C^F_1$ and $C^{\nabla s}_1$ that receive no
contribution from any other piece of a typical density-dependent interaction.
We again set $C^{\Delta s} = 0$, and refit $C^s_1$ (now to 181 $\textrm{MeV fm}^3$) to reproduce the
$^{208}$Pb resonance.  Fig.~\ref{fig:halflives-skop} shows that even these
modifications, which (again) do not alter other predictions, mimic much of the
reduction in half-lives produced by isoscalar pairing. 

These results suggest that the time-odd piece of the functional is much
richer than previously suspected.  The time is ripe for a much more careful
analysis of all these terms.  Our \pnfam\ will allow data from charge-exchange
reactions in deformed nuclei to be included in fits.  A like-particle version would increase the
range of usable data further.  Together with modern optimization techniques,
the efficient calculation of linear response should make for a vast improvement in
our ability to describe beta decay and predict it in important $r$-process
isotopes where measurement is not possible.

\section{Conclusions}\label{sec:conclusions}

We have adapted the finite amplitude method for the computation of beta-decay strength functions and rates in axially-deformed even-even nuclei with
modern Skyrme-like energy-density functionals.  While formally equivalent to
the traditional matrix QRPA, the FAM is far more robust and just as useful as
long as the full set of QRPA energies and transition-matrix elements is
not needed.

To demonstrate the pnFAM's power, we have taken a first look at the effect
of Skyrme's tensor terms on allowed and first-forbidden beta decay in
open-shell isotopes.  We find that the tensor
interaction lowers half-lives in deformed nuclei much like it does in the spherical nuclei studied in
Ref.\ \cite{Minato13}.
Working with the functional \skop, we are able to roughly reproduce measured
rates in a range of nuclei without strong isoscalar pairing and without
spoiling the predictions of the functional in even-even systems.  It is clearly
time to explore time-odd functionals systematically, and we intend to do so
soon.  Finally, beta decay is only one possible application of the
pnFAM.  Neutrino scattering, hadronic charge exchange, and double-beta decay
are three others that come quickly to mind.

\begin{acknowledgments}
We thank Dr.\ Markus Kortelainen, Dr.\ Nobuo Hinohara and Prof.\ Witold
Nazarewicz for useful discussions, and Dr.\ Ewing Lusk for help with
load-balancing in our computations. Support for this work was provided through
the Scientific Discovery through Advanced Computing (SciDAC) program funded by
U.S.\ Department of Energy, Office of Science, Advanced Scientific Computing
Research and Nuclear Physics, under award number DE-SC0008641, ER41896 and by the U.S. Department of Energy Topical Collaboration for Neutrinos and Nucleosynthesis in Hot and Dense Matter, under award number DE-SC0004142.  
Z.~Z.\ acknowledges the support of the TUBITAK-TURKEY, Fellowship No:2219.
We used resources at the National
Energy Research Scientific Computing Center, which is supported by the Office
of Science of the U.S. Department of Energy under Contract No.
DE-AC02-05CH11231.

\end{acknowledgments}

\appendix

\section{Moment of Inertia}\label{app:moi}

The Beliaev formula \cite{Beliaev61} for the moment of inertia in the BCS
approximation is easy to extend to the HFB approximation.  Writing the wave
function of the rotating state to the first order in the rotational speed
$\omega$
\begin{equation}
	|\Psi\rangle = |\textrm{HFB}\rangle + \omega \sum_{\alpha<\beta}
	\frac{\langle\textrm{HFB}| a_\beta a_\alpha J_x
	|\textrm{HFB}\rangle}{E_\alpha + E_\beta} a^\dag_\alpha a^\dag_\beta
	|\textrm{HFB}\rangle \,,
\end{equation}
yields for the moment of inertia
\begin{equation}
\mathcal{I} = \frac{1}{\omega} \langle\Psi| J_x |\Psi\rangle =
\sum_{\alpha\beta} \frac{|\langle\textrm{HFB}|a_\beta a_\alpha J_x
|\textrm{HFB}\rangle|^2}{E_\alpha + E_\beta} \,.
\end{equation}
After applying this transformation to the operator $J_x$ and carrying out the
contractions, we obtain the expression
\begin{equation}\label{eq:moi}
	\mathcal{I} = \sum_{\alpha\beta} \frac{|(U^\dag J_x V^* - V^\dag J_x U^*)_{\alpha\beta}|^2}{E_\alpha + E_\beta} \,.
\end{equation}
The Beliaev formula \cite{Beliaev61}
\begin{equation}
	\mathcal{I}_\textrm{Beliaev} = 2 \sum_{k < k'} \frac{|\langle k|J_x |k'\rangle|^2}{E_k + E_{k'}} (u_k v_{k'} - v_k u_{k'})^2
\end{equation}
applies in the special case that the quasiparticle transformation is diagonal.

\section{First-forbidden Beta Decay}\label{app:forbidden-formulas}

When forbidden operators contribute non-negligibly to beta decay, the
transition strength $B_i$ in \eqref{eq:decayrate} must be replaced by a more
general integrated shape function
\begin{equation}\label{eq:avg-ff-c}
\overline{C}_{J^\pi} = \frac{1}{f(E_0)} \int_1^{W_0} \dd W\, C_{J^\pi}(W) F_0 L_0
pW\left(W_0-W\right)^2.
\end{equation}

Six different multipole operators contribute to non-unique first-forbidden
decay:
\begin{subequations}
\begin{equation}
   \hat{\mathcal{O}}_{ps0} = \frac{\hbar c}{2M_n c^2} \bm \sigma \cdot \bm \nabla \, \tau_-,
\end{equation}
\begin{equation}
   \hat{\mathcal{O}}_p (K) = \frac{\hbar c\,\Theta_K}{2M_n c^2} \nabla_K \,  \tau_-,
\end{equation}
\begin{equation}
   \hat{\mathcal{O}}_r (K) = \frac{\sqrt{3}\,\Theta_K}{R \lambdabar_e} \sqrt{\frac{4\pi}{3}} \, r Y_{1K}(\hat{\vv r}) \, \tau_-,
\end{equation}
and
\begin{equation}
   \hat{\mathcal{O}}_{rsL} (K) = \frac{(-1)^L\,\Theta(K)}{R \lambdabar_e} \sqrt{4\pi} \, r [\vv Y_1 \bm\sigma]_{LK} \, \tau_-,
\end{equation}
\end{subequations}
where $L=0,1,2$.  Here $M_n = 939.0$ MeV$/c^2$ is the nucleon mass, and
$\lambdabar_e= \hbar c / (m_e c^2) = 386.159268$ fm is the (reduced) electron
Compton wavelength.  All operators and resulting quantities are normalized to
the electron mass so that the quantity $\overline{C}_{J^\pi}$ in
Eq.~\eqref{eq:avg-ff-c} is dimensionless \cite{Behrens82}.  The factors
$\Theta_K$ arise from the transformation from intrinsic to laboratory reference
frames \cite{BM98}:
\begin{equation}
   \langle LK | \hat{\mathcal{O}}_{LK} | 00\rangle
   = \Theta_K \langle K | \hat{\mathcal{O}}_{LK} | 0 \rangle,
\end{equation}
where
\begin{equation}
   \Theta_K = \begin{cases}1, & K = 0 \\ \sqrt{2}, & K > 0 \end{cases}.
\end{equation}

The shape factors for the non-unique forbidden decay are worked out e.g.\ in
Ref.~\cite{Suhonen93}.  Expressing the squared matrix elements and the
interference terms in terms of residues and replacing the kinematic parts of integrands
by polynomial expressions that closely approximate them along a portion of the
real axis (see main text) one can write first-forbidden shape functions in the
form
\begin{equation}
\overline{C}_{J^\pi} \approx \frac{1}{2\pi i} \sum_i \oint_C \dd\omega\, P_i
(\omega) R_i(J^\pi; \omega) \, ,
\end{equation}
where the $R_i$ are linear combinations of functions $S(F;\omega)$ and
$\chi(F,G;\omega)$ is defined in Eqs.\ \eqref{eq:strfunc} and \eqref{eq:crossterm}.  For $J^\pi = 0^-$ we have
\begin{subequations}
\begin{equation}
R_1(0^-,\omega) = -\frac{2}{3} g_A^2 \big(X_+ S(\hat{\mathcal{O}}_{rs0};
\omega) + \chi(\hat{\mathcal{O}}_{rs0}, \hat{\mathcal{O}}_{ps0}; \omega) \big)
\end{equation}
and
\begin{equation}\begin{split}
   R_2(0^-&,\omega) = g_A^2 \Bigg[ \left(X_+^2 + \frac{1}{9} \right) S(\hat{\mathcal{O}}_{rs0}; \omega) \\ 
   &+ S(\hat{\mathcal{O}}_{ps0}; \omega) + 2X_+ \chi(\hat{\mathcal{O}}_{rs0},
   \hat{\mathcal{O}}_{ps0}; \omega) \Bigg] \,,
\end{split}\end{equation}
\end{subequations}
for $J^\pi = 1^-$ we have
\begin{subequations}
\begin{widetext}
\begin{equation}\begin{split}
   R_1(1^-,\omega) =& -\frac{2}{9} \Big[ X_+ S(\opr;\omega) -2g_A^2 X_- S(\oprsone;\omega)
   - g_A \sqrt{2} (X_+ - X_-) \chi(\opr,\oprsone;\omega) \\
   &- \sqrt{3} \chi(\opp,\opr;\omega) + g_A \sqrt{6} \chi(\opp,\oprsone;\omega) \Big],
\end{split}\end{equation}
\begin{equation}\begin{split}
   R_2(1^-,\omega) =& S(\opp;\omega) + \frac{1}{3} X_+^2 S(\opr;\omega)
   + \frac{2}{3} g_A^2 X_-^2 S(\oprsone;\omega)
   - \frac{8}{27} \big( g_A^2 S(\oprsone;\omega)
   - \frac{g_A}{\sqrt{2}} \chi(\opr,\oprsone;\omega) \big)\gamma_1 \\
   &+\frac{1}{27} \big( S(\opr;\omega) + 2g_A^2 S(\oprsone;\omega)
   - 2\sqrt{2} g_A \chi(\opr,\oprsone;\omega) \big) \\
   &+\sqrt{\frac{2}{3}} \Big( - 2g_A X_- \chi(\opp,\oprsone;\omega) - \sqrt{2} X_+ \chi(\opp,\opr;\omega)
   + \frac{2}{\sqrt{3}} g_A X_- X_+ \chi(\opr,\oprsone;\omega) \Big) \,,
\end{split}\end{equation}
\begin{equation}\begin{split}
   R_3&(1^-,\omega) = \frac{4}{3} \Big[ -\frac{\sqrt{2}}{3} g_A X_+ \chi(\opr,\oprsone;\omega)
   - \frac{2}{3} g_A^2 X_- S(\oprsone;\omega) 
   + \sqrt{\frac{2}{3}} g_A \chi(\opp,\oprsone;\omega) \Big],
\end{split}\end{equation}
\end{widetext}
\begin{equation}
   R_4(1^-,\omega) = \frac{8}{27} g_A^2 S(\oprsone;\omega) \,,
\end{equation}
\begin{equation}\begin{split}
   R_5(1^-,\omega) &= \frac{1}{27} \Big[ 2 S(\opr;\omega) + g_A^2 S(\oprsone;\omega) \\
   &-2\sqrt{2} g_A \chi(\opr,\oprsone;\omega) \Big] ,
\end{split}\end{equation}
\begin{equation}\begin{split}
   R_6(1^-,\omega) &= \frac{1}{27} \Big[ 2 S(\opr;\omega) + g_A^2 S(\oprsone;\omega) \\
   &+2\sqrt{2} g_A \chi(\opr,\oprsone;\omega) \Big] \,,
\end{split}\end{equation}
\end{subequations}
and finally, for $J^\pi = 2^-$ have 
\begin{equation}
R_5(2^-,\omega) = R_6(2^-,\omega) = \frac{1}{9} g_A^2
S(\hat{\mathcal{O}}_{rs2}; \omega) \,.
\end{equation}
We have used the shorthand 
\begin{equation}
X_\pm = \left( \frac{W_0}{3} \pm \frac{\alpha Z}{2R} \right) \,,
\end{equation}
where $\alpha$ is the fine-structure constant and $R$ is the nuclear radius.
The polynomials $P_k (\omega)$ are fitted to the various integrated kinematical
factors so that $P_i (\omega) \approx G_i((\omega_\text{max} - \omega)/(m_e
c^2) + 1)$, with 
\begin{equation}
\label{eq:phasespaceints}
G_i (W_0) = \int_1^{W_0} \dd W \, g_i F_0 L_0 p (W_0-W)^2 \,
\end{equation}
in the interval $\omega \in [0,\omega_\text{max}]$ (within our contour).  Here
\begin{equation}\begin{split}
   g_1 &= \gamma_1, \, g_2 = W, \, g_3 = W^2 \,, \\
   g_4 &= W^3, \, g_5 = W(W_0-W)^2 \,,\textrm{ and} \\
   g_6 &= \lambda_2 W(W^2-1) \,,
\end{split}\end{equation}
where the function $\lambda_k$ is
\begin{equation}
   \lambda_k = \frac{(k+\gamma_k)F_{k-1}}{k(1+\gamma_1)F_0} \,.
\end{equation}

The polynomial $P_2$ above is the same one that enters the computation of
allowed decay.  (We called it $f_\textrm{poly}$ in the main text.)

% The bibliography will be pasted below
%\bibliography{refs}
%

\end{document}